\RequirePackage{fix-cm}
\documentclass[final, twocolumn]{svjour3}
\usepackage{graphicx}
\usepackage{amsmath}
\usepackage{verbatim}
\usepackage[numbers]{natbib}
\usepackage{hyperref}
\usepackage{fullpage}

\journalname{Journal of Materials Science}

\begin{document}
\title{A QM/MM Approach for the Study of Monolayer Protected Gold Clusters}

\author{Sandipan Banerjee         	    \and
             John A. Montgomery, Jr.      \and
             Jos\'e A. Gasc\'on      	      }

\institute{Sandipan Banerjee \and John A. Montgomery, Jr.  \at
              Department of Physics, University of Connecticut \\Storrs, CT 06269, USA. \\
             \email{banerjee@phys.uconn.edu \\ johnm@phys.uconn.edu}\\
          \and
           Jos\'e A. Gasc\'on \at
           Department of Chemistry, University of Connecticut \\Storrs, CT 06269, USA. \\
           Tel.: +1--860--486--0591\\
	  \email{jose.gascon@uconn.edu}\\
}

\date{Received: date / Accepted: date}

\maketitle

\section*{Abstract}
We report the development and implementation of hybrid methods that combine Quantum Mechanics (QM) with
Molecular Mechanics (MM) to theoretically characterize thiolated gold clusters. We use, as training
systems, structures such as Au$_{25}$(SCH$_2$-R)$_{18}$ and Au$_{38}$(SCH$_2$-R)$_{24}$, which can be readily compared with recent
crystallographic data. We envision that such an approach will lead to an accurate description of key
structural and electronic signatures at a fraction of the cost of a full quantum chemical treatment. As an example, we demonstrate that calculations of the $^1$H and $^{13}$C NMR shielding constants with our proposed
QM/MM model maintain the qualitative features of a full DFT calculation, with an order-of-magnitude increase
in computational efficiency.



\section{Introduction}

\label{intro}
It is well known that gold nanoparticles are capable of catalyzing a number of fundamental chemical reactions \cite{Haruta2005,Corma2006,Hutchings2006}. Tunable reactions in metallic nanoparticles can be obtained by encapsulation of nanocrystals in thiolated monolayers \cite{Templeton2000,Zhong2001} known as monolayer protected clusters (MPCs). Exploiting MPCs for applications in the area of catalysis requires a proper structural and dynamical characterization of the physical properties of the protecting layer. It was not until recently, that such requisite has been met by the total structure determination of a number of thiolated gold clusters \cite{RogerAu102,MurrayAu25,JinAu38}. Such structures are characterized by surface gold atoms in the so called ``staple'' (--S--Au--S--) or ``V-shape'' motifs (--S--Au--S--Au--S--), as opposed to commonly assumed structures in which thiolates only passivate a high symmetry gold cluster. Along these experimental studies, Density Functional Theory has been crucial to interpret structural data under the light of NMR spectroscopy \cite{Gascon-Maran2010,Gascon-Maran2011}, optical absorption \cite{Aikens2009}, and electrochemical experiments \cite{Parker2010}. Due to the size of these clusters, augmented by thiolated ligands, theoretical characterization (e.g calculation of spectra) requires an enormous amount of computational resources (both in time and memory). Intrinsic limitations of current Quantum Chemistry software places these types of calculations at the limits of computational tractability. This is clearly the case if, in addition to quantum detail, finite temperature simulations are required. In fact, a number of studies have shown a marked dependence of optical and electrochemical properties on temperature \cite{Devadas2011,temperature1,temperature2}. Thus, to retain quantum detail and incorporate finite temperature motions it would certainly be advantageous to develop approaches that can combine these features with a reasonable computational cost.\\

Empirical potentials have been used in the past for gold clusters to search structural minima followed by quantum mechanical refinement \cite{Soler1999}, but little has been done in the context of MPC's, especially considering the recent re-evaluation of the sulfur-gold binding motifs. Another approach would be to develop hybrid methods such as QM/MM, which were originally developed in the context of enzyme reactions \cite{Warshel1976}. In this work we develop MM and QM/MM models, explore different implementations, and evaluate their ability to reproduce experimental data. We envision that such approaches will be of value to enable, in a practical manner, sampling of a large number of monolayer conformations. In particular, such procedures will be relevant to describe highly interacting oligopeptide ligands  \cite{Fabris2006,Gascon-Maran2010}. In addition, partitioning the system into a QM and MM region will lead to several advantages. Apart from the obvious computational benefits, such method opens up a wide range of possibilities for certain studies of similar systems for which a full QM description is impractical and often unnecessary. This new approach also opens the possibility of focusing strictly on the electronic structure of the monolayer alone (or of the gold cluster and a fraction of the monolayer) which, for instance, can be advantageous to obtain insight into catalytic properties of these MPC's\cite{Mita2011}. Thus, the goal of this study is to develop a first generation of hybrid methods that can both decrease the computational cost and, at the same time, reproduce the intricacies of the various gold-sulfur surface motifs. \\

 We use [Au$_{25}$(SCH$_3$)$_{18}$]$^{-}$ as the central prototype model and extend our theory to explain structural and chemical properties of larger MPC's. We begin by determining a minimum set of parameters and functional forms that would be sufficient to describe a thiol-protected gold cluster via a Molecular Mechanics (MM) force field. We then use these parameters to develop a hybrid QM/MM model which treats regions of chemical interest at a DFT level of theory, keeping all other atoms in the cluster at the MM level. \\

\section{Computational Methods}

We propose the use of hybrid QM/MM models to study physical and chemical properties of monolayer protected gold clusters. Au-SCH$_3$, Au-(SCH$_3$)-Au, and [Au$_{25}$\-(SCH$_3$)$_{18}$]$^{-}$ are used as the training systems. In the following sections, we present the parameters and functions necessary to describe the structure in a molecular mechanics (MM) framework. We continue our discussion by introducing QM/MM hybrid models to predict or refine structures and NMR properties of similar gold clusters. All QM/MM calculations were performed with the Gaussian 09 suite \cite{g09} using the two-layer ONIOM scheme \cite{Dapprich1999,Vreven2006}. In this scheme, the entire system is divided into two regions ($X$ and $Y$). The QM/MM energy is obtained via an extrapolation of three independent calculations:

\begin{equation}
E = E(QM)_X + E(MM)_{X+Y} - E(MM)_X \\
\label{eq:oniom}
\end{equation}

\noindent where $E(QM)_X$ is the energy of region $X$ at the QM level, $E(MM)_{X+Y}$ is the energy of the entire system ($X$ and $Y$) at the MM level, and $E(MM)_X$ is the energy of $X$ at the MM level. In the so called ``electronic embedding'' (EE) approach, electrostatic interactions between $X$ and $Y$ are included in each of the terms of the right hand side of Eq.~\ref{eq:oniom}, so that electrostatic interactions are canceled out at the MM level, but remain at the QM level. In the ``mechanical embedding'' (ME) approach, electrostatic interactions between region $X$ and $Y$ are only included at the MM level (the last two terms of Eq.~\ref{eq:oniom}). In both EE and ME, Van der Waals (VDW) interactions between $X$ and $Y$ are considered at the MM level (in the term $E(MM)_{X+Y}$). All calculations presented here make use of the EE approximation.  \\

\subsection{Molecular Mechanics Model}
\label{sec:MM}
\begin{figure}
\begin{center}
  \includegraphics[clip, width=1.00\columnwidth]{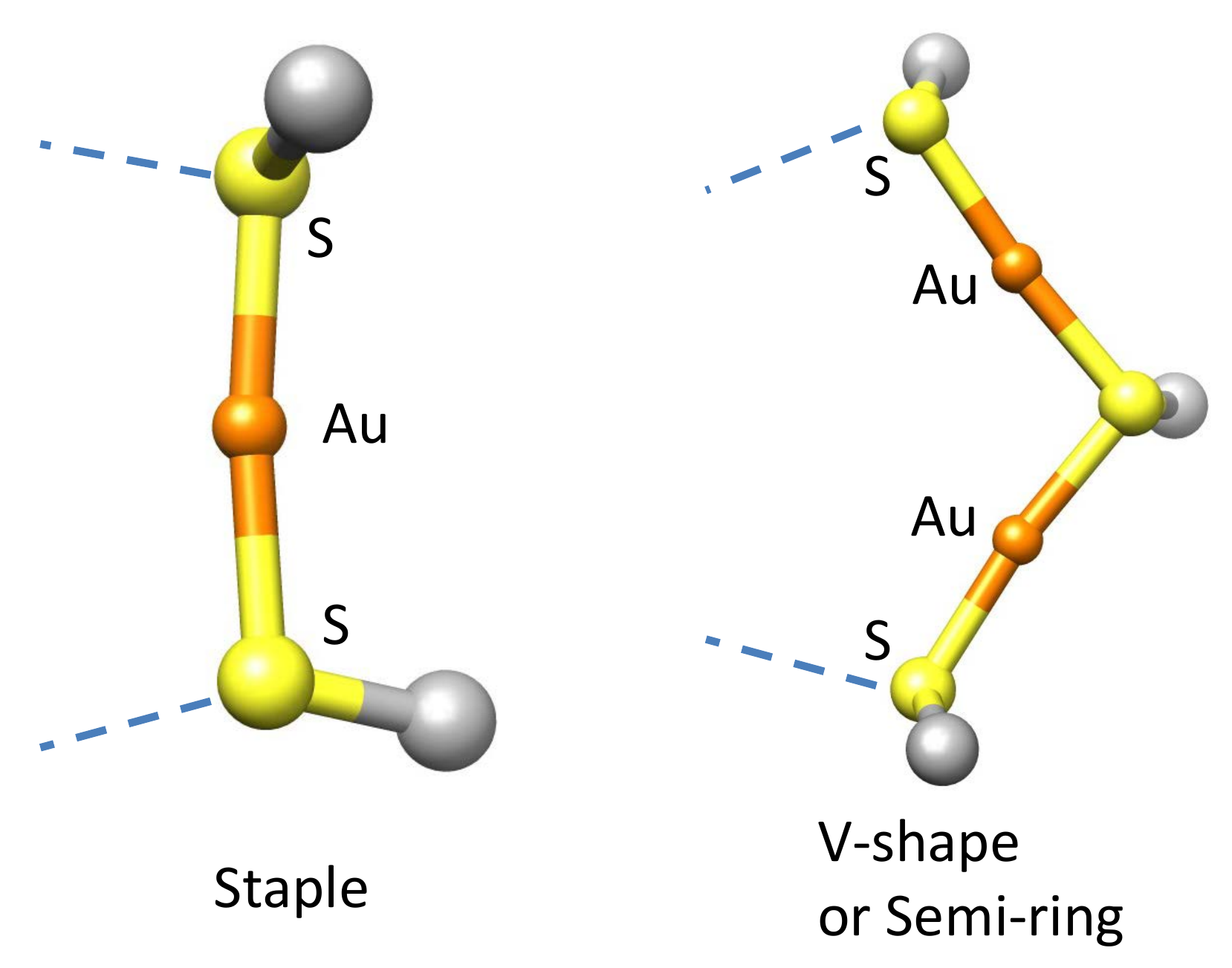}
\caption{[COLOR ONLINE] Possible binding patterns between gold and thiolated ligands}
\label{fig:binding}
\end{center}
\end{figure}

Construction of QM/MM models was made under the assumption that even a pure MM force field description of the thiolated clusters should reproduce the experimental structural data, at least in a qualitative and semi-quantitative manner.  To this end, we sought to develop a force field with a minimal set of parameters and functional forms. Derivation of the force field was guided by the notion that these clusters present two well defined domains. Domain {\bf 1} contains the inner core (i.e. 13 Au atoms for Au$_{25}$(SCH$_2$-R)$_{18}$ and 23 Au atoms for Au$_{38}$(SCH$_2$-R)$_{24}$) forming high symmetry packed structures. For this domain we defined a 6-12 Lennard-Jones potential as implemented in the Amber force field \cite{amber}. Domain {\bf 2} contains the thiolated ligands and all Au atoms involved in the V-shape or staple motifs (Fig. \ref{fig:binding}). All gold atoms in this domain are assigned the same force field atom type (AuS). As starting values we took those reported in the UFF force field of Rapp\'e {\it et al} \cite{UFF}. These parameters were modified systematically and stepwise until the geometry of [Au$_{25}$(SCH$_3$)$_{18}$]$^{-1}$ was reproduced within a tolerance of 0.2 \AA \  in the root mean square deviation. Gold atoms in {\bf 1} are separated from those that are only bonded to other core atoms (type Au) and those that are connected to core atoms and sulfur atoms (type AuC). Although the VDW parameters are identical for these two types, this distinction is required for the definition of bonded parameters (vide infra). VDW parameters are reported in Table \ref{tab:VDW}. \\

\begin{table}
\caption{6-12 Lennard Jones parameters for a general MPC. Partial charges correspond to thiol in [Au$_{25}$(SCH$_3$)$_{18}$]$^{-1}$}
\label{tab:VDW}
\centering
\begin{tabular}{llllr}
\hline\noalign{\smallskip}
Atom & Atom-type & $\sigma$ & $\epsilon$  & charge \\
\noalign{\smallskip}\hline\noalign{\smallskip}
Au	& Au         & 1.5000 & 1.0000 & 0.00000\\
Au 	& AuC       &   1.5000 & 1.0000 & 0.00000\\
Au 	& AuS       &  1.9000 & 1.0000 & 0.00000\\
S 	& SS          &  2.0000 & 0.2500 & -0.22417\\
S 	& SC 	&  2.0000 & 0.2500 & -0.22417\\
C 	& CT	& 1.9080 & 0.1094 & -0.15441\\
H	& H1		& 1.3870 & 0.0157 & 0.10767\\
\noalign{\smallskip}\hline
\end{tabular}
\end{table}

Electrostatic interactions between the ligands and gold atoms in domain {\bf 1} are assumed to be negligible. Support for this assumption is given by a Bader charge analysis \cite{Bader1990,Henkelman2006} of [Au$_{25}$(SCH$_3$)$_{18}$]$^{-1}$ which shows that charges in domain {\bf 1} are very small ($\approx -0.04e$ for the central core atom and $\approx 0.02e$ for the other core atoms), adding up to a total charge of 0.15$e$. Bader charges for AuS atoms are also small ($\approx 0.04e$). Thus, to facilitate transferability among different core sizes, the charges on all gold atoms are assumed to be zero. Charges on the thiol ligands are taken from an Electrostatic Potential (ESP) analysis (as implemented in Gaussian 09) for the neutral molecule Au$_2$-[SCH$_3$]$_2$. This analysis provides the following charges: $q_{\textrm{S}}=-0.23605$, $q_{\textrm{C}}=-0.16628$, and $q_{\textrm{H}}=0.09580$. To ensure that the total MM charge adds up to the total charge of the cluster ($Q$), a simple correction $\delta q$ is applied to each charge via the formula $\delta q = (Q/M - (q_{\textrm{S}} + q_{\textrm{C}} + 3q_{\textrm{H}}))/5$, where $M$ is the number of thiolated ligands. Table~\ref{tab:VDW} reports these charges for $Q=-1$ and $M=18$, which uses the correction $\delta q = 0.01188$.

Bonded parameters (stretching, bending, torsion) involving X-Au-S-X were derived by fitting the force constants in all functional forms to match vibrational frequencies of selected modes of vibration for the mo\-le\-cules Au-SCH$_3$ and (Au-SCH$_3$-Au). All other parameters intrinsic to SCH$_3$ were taken from the Amber force field without modification. Tables \ref{tab:STC}, \ref{tab:BEN}, and \ref{tab:TOR} report all bonded parameters involving the newly defined atom types and the existing types in the Amber force field.

\begin{table}
\caption{Stretching parameters corresponding to the force field formula $(1/2)k_{e}(r-r_{eq})^2$.}
\label{tab:STC}
\centering
\begin{tabular}{llll}
\hline\noalign{\smallskip}
At-type1 & At-type2 & k$_r$ & r$_{eq}$  \\
\noalign{\smallskip}\hline\noalign{\smallskip}
AuC & SC	& 150.00 & 2.48 \\
AuS	& SC	 & 150.00 & 2.38 \\
AuS	& SS	 & 150.00 & 2.38 \\
SC	& CT & 237.00 & 1.84 \\
SS	& CT & 237.00 & 1.84  \\
CT	& H1	 & 340.00 & 1.09 \\
CA	& H1 & 340.00 & 1.09 \\

\noalign{\smallskip}\hline
\end{tabular}
\end{table}

\begin{table}
\caption{Bending parameters corresponding to the force field formula $(1/2)k_{\theta}(\theta-\theta_{eq})^2$.}
\label{tab:BEN}
\centering
\begin{tabular}{lllll}
\hline\noalign{\smallskip}
At-type1 & At-type2 & At-type3 & k$_\theta$ & $\theta_{eq}$  \\
\noalign{\smallskip}\hline\noalign{\smallskip}
AuC	& SC	 & 	CT  	 & 35.00  & 105.00  \\
AuS	& SC & 	CT  	 & 35.00  & 105.00  \\
SC	& CT &	H1  	 & 50.00  & 109.50  \\
H1	& CT &	H1  	 & 35.00  & 109.50  \\
SC	& AuS &	SS  	 & 20.00  & 180.00  \\
AuS	& SS &	CT  	 & 35.00  & 105.00  \\
AuS	& SS	& 	AuS 	 & 20.00  & 100.00  \\
AuC & SC &	AuS 	 & 20.00   & 91.00  \\
SS	& CT &	H1  	 & 50.00  & 109.50 \\
SC	& CT &	CT  	 & 50.00  & 111.50  \\
SS	& CT &	CT  	 & 50.00  & 111.50  \\
H1	& CT &	CA  	 & 35.00  & 110.50  \\
CA	& CA &	H1  	 & 35.00  & 120.00  \\

\noalign{\smallskip}\hline
\end{tabular}
\end{table}

\begin{table*}
\caption{Amber torsional parameters corresponding to the force field formula $\sum_{i=1}^{4} M_i[1 + \cos(i\theta - O_i(i+4))]/N_p$.}
\label{tab:TOR}
\centering
\begin{tabular}{lllllllllllll}
\hline\noalign{\smallskip}
At-type1 & At-type2 & At-type3 & At-type4 & $O_1$ & $O_2$ & $O_3$ & $O_4$ & $M_1$ & $M_2$ & $M_3$ & $M_4$ & $N_p$\\
\noalign{\smallskip}\hline\noalign{\smallskip}
 AuC & SC & CT & H1 & 0 & 0 & 0 & 0 &  0.00 &  0.00 &  0.16 &  0.00 & 1.00 \\
AuC & SC & CA & H1 & 0 & 0 & 0 & 0 &  0.00 &  0.00 &  0.16 &  0.00 & 1.00 \\
 AuS & SC & CT & H1 & 0 & 0 & 0 & 0  & 0.00 &  0.00 &  0.16 &  0.00 & 1.00 \\
 AuS & SC & CA & H1 & 0 & 0 & 0 & 0 &  0.00 &  0.00 &  0.16 &  0.00 & 1.00 \\
 AuS & SS & CT  & H1 & 0 & 0 & 0 & 0  & 0.00  & 0.00 &  0.16 &  0.00 & 1.00 \\
AuS & SS & CA & H1 & 0 & 0 & 0 & 0 &  0.00 &  0.00 &  0.16 &  0.00 & 1.00 \\
SC  & AuS & SC & CT & 0 & 0 & 0 & 0 &  0.00 &  0.00 &  0.00 &  0.00 & 1.00 \\
SC  & AuS & SC  & CA & 0 & 0 & 0 & 0 &  0.00 &  0.00 &  0.00 &  0.00 & 1.00 \\
SC  & AuS & SS &  CT  & 0 & 0 & 0 & 0 &  0.00 &  0.00 &  0.00 &  0.00 & 1.00 \\
SC  & AuS & SS & CA & 0 & 0 & 0 & 0 &  0.00 &  0.00 &  0.00 &  0.00 & 1.00 \\
AuS & SS & AuS & SC & 0 & 0 & 0 & 0 &  0.00 &  0.00 &  0.00 &  0.00 & 1.00 \\
AuC & SC & AuS & SS & 0 & 0 & 0 & 0 &  0.00 &  0.00 &  0.00 &  0.00 & 1.00 \\
  CT  & SC & AuS & SS & 0 & 0 & 0 & 0 &  0.00 &  0.00 &  0.00  & 0.00 & 1.00 \\
CA & SC & AuS & SS & 0 & 0 & 0 & 0 &  0.00 &  0.00 &  0.00 &  0.00 & 1.00 \\
\noalign{\smallskip}\hline
\end{tabular}
\end{table*}

\subsection{QM/MM Models}
\subsubsection{Au$_{25}$}

\begin{figure}
\begin{center}
  \includegraphics[clip, width=1.00\columnwidth]{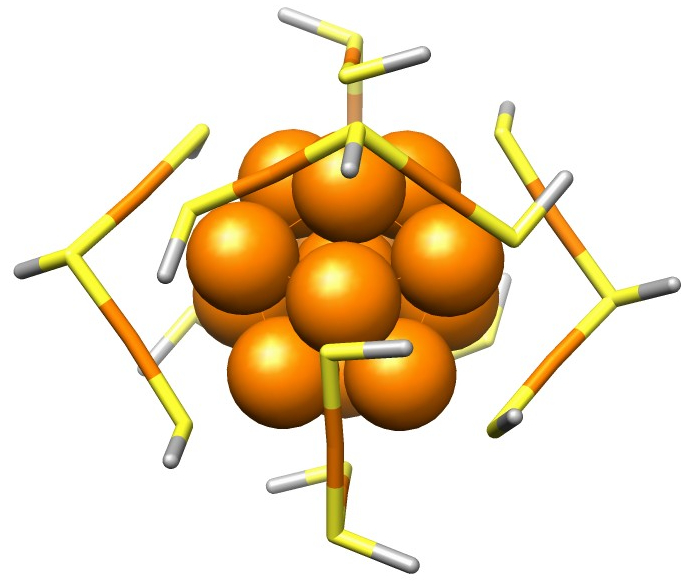}
\caption{[COLOR ONLINE] Structure of [Au$_{25}$(SCH$_3$)$_{18}$]$^{-}$, with Au atoms shown in orange, S in yellow, C in gray. H atoms are not shown for clarity. The Au$_{13}$ core (MM region) is shown with spheres while all other atoms (QM region) are shown with sticks.}
\label{fig:gold25}
\end{center}
\end{figure}

One of the major objectives in developing a full MM force field is to use it as the underlying potential in hybrid QM/MM models. We benchmark our model by comparing structural descriptors in [Au$_{25}$(SCH$_3$)$_{18}$]$^{-}$. The crystal structure of [N(C$_8$H$_{17}$)$_4$ ]$^+$[{Au$_{25}$(SCH$_2$CH$_2$Ph)$_{18}$}]$^{-1}$ was determined recently by Heaven {\it et al.} \cite{MurrayAu25}, which was followed by several other experimental investigations \cite {Zhu2008} and theoretical \cite{Akola2008} calculations using density functional theory (DFT) methods. The initial geometry for our QM/MM calculations is derived from that crystal structure by replacing the ethylphenyl groups with methyl groups. We adjust the resulting [Au$_{25}$(SCH$_3$)$_{18}$]$^{-}$ structure to obtain a symmetric structure belonging to the point group C$_i$. \\

The [Au$_{25}$(SCH$_3$)$_{18}$]$^{-}$ structure consists of an icosahedral Au$_{13}$ core, protected by six V-shape motifs in an approximate octahedral arrangement. Each of the S atoms are connected to the organic ligand (CH$_3$), forming a monolayer protected cluster. We propose a QM/MM model in which the icosahedral Au$_{13}$ core is treated via MM whereas the V-shape motifs with the ligands are treated via QM methods(see figure \ref{fig:gold25}). These two regions interact with each other via the ONIOM extrapolation scheme with electronic embedding as described in Eq.~\ref{eq:oniom}.


\begin{figure}
\begin{center}
  \includegraphics[clip, width=1.00\columnwidth]{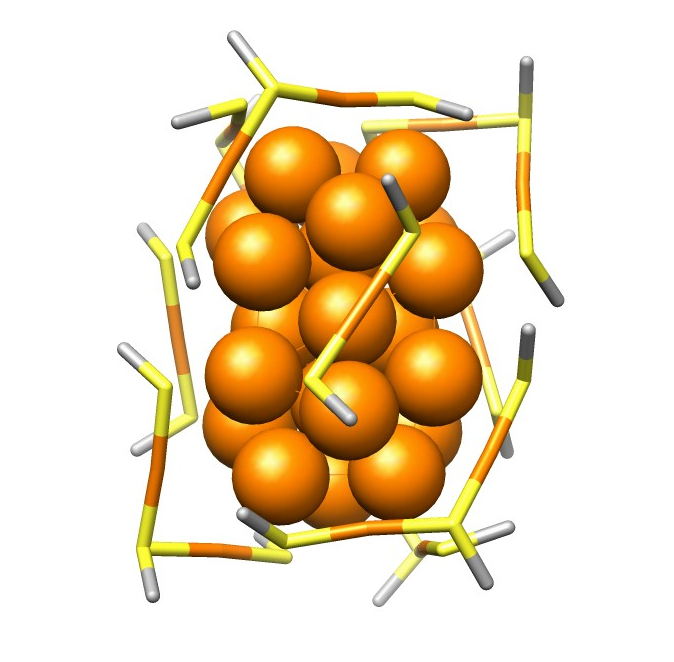}
\caption{[COLOR ONLINE] Structure of Au$_{38}$(SCH$_3$)$_{24}$. The Au$_{23}$ core (MM region) is shown with spheres while all other atoms (QM region) are shown with sticks.}
\label{fig:gold38}
\end{center}
\end{figure}

Our choice of the partition for [Au$_{25}$(SCH$_3$)$_{18}$]$^{-}$ cluster was fairly straight-forward, making use of the intuitive definition of domains presented in the previous section. The icosahedral Au$_{13}$ core is very stable and can be accurately described by a Lennard-Jones potential. Each of the gold atoms in the core, except the central one, is bonded to a sulphur atom forming a V-shape motif. A ``cut'' is defined between atoms types AuC and SC. These bond cuts are implemented in ONIOM by defining a ``link-atom" to satisfy the valence and avoid broken bonds. The default choice of such a link-atom is typically a hydrogen atom. However for our purposes, the best choice of link-atom turned out to be another Au atom, thus forming Au(link)--S bonds in place of Au(core)--S bonds. Since the core has a small charge ($\approx 0.15e$), and since the QM region needs to be assigned an integer charge, we assume a --1 charge solely distributed among the atoms of domain {\bf 2} (QM region), thus making the entire structure an anion. The force field (based on Amber) described previously was used for the MM region. DFT was used to describe the remaining 102 atoms in domain {\bf 2}. We used the LANL2DZ psuedopotential and basis sets as defined in Gaussian 09 with the BLYP functional. For computational efficiency, we use density fitting with the W06 fitting basis \cite{weigend2006}. It is evident that for this type of partition the gain in computational time, with respect to a full QM calculation, is minimal, since the QM region has effectively 24 gold atoms. Nevertheless, such partitioning scheme represents a proof of concept and will clearly become computationally more efficient for larger clusters and for cases where a reduced number of ligands are described quantum mechanically.   \\

\subsubsection{Au$_{38}$}

As a second example, we use our QM/MM model to optimize the structure of Au$_{38}$(SCH$_3$)$_{24}$. The X-ray structure was determined recently \cite{Qian2010} along with a full DFT comparison \cite{Acevedo2010}. The Au$_{38}$(SCH$_3$)$_{24}$ structure has a bi-icosahedral Au$_{23}$ core, consisting of two icosahedral Au$_{13}$ units joined by three shared Au atoms at the center. This core is protected by six long [(Au)$_2$(SCH)$_3$] semi-ring (V-shape motifs) and three short [Au(SCH3)$_2$] semi-rings (staple motifs) (see figure \ref{fig:gold38}). The entire structure has D$_3$ symmetry, the largest Abelian point group being C$_2$.\\

By analogy with our partitioning of the [Au$_{25}$ (SCH$_3$)$_{18}$]$^{-}$ system, we describe the bi-icosahedral Au$_{23}$ core in Au$_{38}$(SCH$_3$)$_{24}$ by our modified Amber force field. The outer shell comprising the V-shape and staple motifs was described quantum mechanically using DFT. As before, we use the BLYP pure functional with LANL2DZ and density fitting basis set W06. Charge of the QM region was assumed to be zero.

\subsubsection{Semi-rings}

\begin{table*}
\caption{Structure validation for different QM/MM models. All distances are reported in Angstroms.}
\label{tab:STR}
\centering
\begin{tabular}{llllll}
\hline\noalign{\smallskip}
Cluster & $d_{21}$ & $(d_{21})_{XRAY}$ & $d_{rms}$ & $(d_{rms})_{XRAY}$ & $\delta_S$ (\%)  \\
\noalign{\smallskip}\hline\noalign{\smallskip} \\
\mbox{[Au$_{25}$(SCH$_3$)$_{18}$]$^{-}$} & 2.84 & 2.87 & 2.73 & 2.76 & 1.09 \\ \\
\mbox{[Au$_{38}$(SCH$_3$)$_{24}$]} & 2.09 & 2.02 & 1.71 & 1.76 & 2.84 \\ \\
\mbox{[Au$_{25}$(SCH$_3$)$_{12}$(SCH$_2$CH$_2$Ph)$_6$]$^{-}$} & 2.87 & 2.87 & 2.72 & 2.76 & 1.45 \\ \\
\noalign{\smallskip}\hline
\end{tabular}
\end{table*}

\begin{table}
\caption{ $^{13}$C NMR chemical shifts $\delta$ (in ppm) for [Au$_{25}$(SCH$_3$)$_{12}$(SCH$_2$CH$_2$Ph)$_6$]$^{-}$ cluster. Reported are the calculated values within the QM/MM approximation, full DFT calculation, and experimental values.}
\label{tab:NMRC}
\centering
\begin{tabular}{llll}
\\ \hline\noalign{\smallskip}
Atom & QM/MM & DFT  & Experiment \cite{Gascon-Maran2011}\\
\noalign{\smallskip}\hline\noalign{\smallskip}
$\alpha$--C$_{inner}$ & 50.75 & 55.80 & --- \\
$\alpha$--C$_{outer}$ & 48.72 & 49.74 &  35.97 \\ \\
$\beta$--C$_{inner}$ & 45.11 & 44.26  & --- \\
$\beta$--C$_{outer}$ & 44.07 & 43.47 & 42.18 \\ \\
i--C$_{inner}$ & 137.35 & 141.05 & 141.88 \\
i--C$_{outer}$ & 137.43 & 140.42 & 140.88 \\ \\
m--C$_{inner}$ & 125.48  & 125.00 & 128.52\\
m--C$_{outer}$ & 125.61 & 124.94 & 128.63 \\ \\
o--C$_{inner}$ & 125.29 & 126.49 & 129.85 \\
o--C$_{outer}$ & 125.37 & 125.42 & 129.30 \\ \\
p--C$_{inner}$ & 123.99 & 122.45 & 126.27 \\
p--C$_{outer}$ & 123.96 & 122.35 & 126.49 \\ \\
\hline
MUE (in ppm) & 3.50 & 3.23 & \\
\noalign{\smallskip}\hline
\end{tabular}
\end{table}

\begin{figure}
\begin{center}
  \includegraphics[clip, width=1.00\columnwidth]{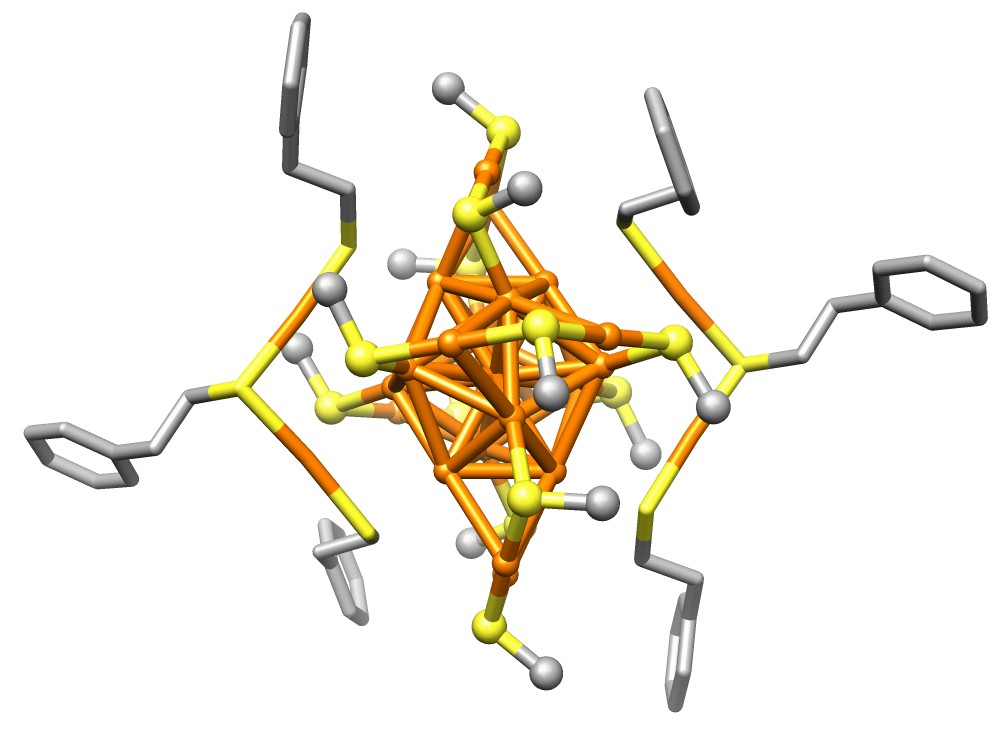}
\caption{Hybrid structure [Au$_{25}$(SCH$_3$)$_{12}$(SCH$_2$CH$_2$ Ph)$_6$]$^{-}$. The two semi-rings, belonging to the QM region, have ethyl-phenyls as ligands (shown by sticks). The remaining four semi-rings have methyls, which together with the Au$_{13}$ core belong to the MM region (shown in balls \& sticks).}
\label{fig:sr}
\end{center}
\end{figure}

We show in our final example, that one can use the same prescription proposed earlier to isolate a portion of the monolayer for a QM description, while retaining an MM description of the remainder. For this purpose, we took the [Au$_{25}$(SCH$_3$)$_{18}$]$^{-}$ structure and changed the ligands in two of the semi-rings from methyl to ethylphenyl, thus forming [Au$_{25}$(SCH$_3$)$_{12}$(SCH$_2$CH$_2$Ph)$_6$]$^{-}$ (see figure \ref{fig:sr}). The QM region contains two semi-rings on opposite sides of the structure and the C$_i$ symmetry of the original structure is preserved.
Again the Au$_{13}$ core and remaining (Au)$_{2}$(SCH$_3$)$_3$ semi-rings are in the MM region. We optimized the geometry and calculated NMR properties of $^{13}$C and $^1$H.  Results are reported in the following section.

\section{Results}

We report in Table \ref{tab:STR} an analysis of the optimized structure for the different clusters described in the previous section.  To assess the accuracy of our structure, we define the parameters $d_{21}$ and $d_{rms}$ as the distance between two neighboring Au atoms in the core and the ``root mean square'' distance of all Au atoms from the central atom in the core, respectively. We define the percent error $\delta_S$ as

\begin{displaymath}
\delta_S = \frac{|(d_{rms})_{XRAY} -  (d_{rms})_{MODEL}|}{(d_{rms})_{XRAY}} \times 100
\end{displaymath}


We note that the maximum error in structure optimization using our QM/MM models is $\sim$ 3 \%, which is remarkably good. We then used the optimized structures to calculate NMR properties of $^{13}$C and $^1$H, which we present in Tables \ref{tab:NMRC} and \ref{tab:NMRH}. Guided by the benchmark studies of Cheeseman {\it et al.} \cite{Cheeseman1996}, our calculations of NMR chemical shifts are performed with the B3LYP hybrid functional, using the LANL2DZ basis set for Au, and the 6--31G(d,p) for S, C and H. We see that the NMR isotropic shieldings obtained using the QM/MM model compare very well to that found using full DFT on the same [Au$_{25}$(SCH$_3$)$_{12}$(SCH$_2$CH$_2$Ph)$_6$]$^{-}$ structure. We also calculate the mean unsigned error (MUE) to compare theory with NMR experiments reported recently \cite{Gascon-Maran2011}. The mean unsigned error is defined as

\begin{displaymath}
MUE = \frac{1}{n_i}\sum_{i}(|\delta_{model}^i- \delta_{experiment}^i|)
\end{displaymath}

\noindent where $n_i$ is the total number of different types of C and H atoms, and $\delta$ are the NMR chemical shifts. The maximum MUE is 3.5 ppm for  $^{13}$C and 0.18 ppm for $^1$H NMR. Thus we see that both $^{13}$C and $^1$H agree remarkably well with the known experimental results. Although small, we believe that the discrepancies arise from the fact that the experimental structure has six semi-rings with CH$_2$CH$_2$Ph as the functional group for all ligands whereas the QM/MM and full DFT models have two semi-rings with CH$_2$CH$_2$Ph as ligands and the rest replaced with CH$_3$, so we are neglecting some of the electrostatic interactions that arise due to the difference in the structure. However, the error seems to be small in comparison to the huge gain in computational efficiency. More precisely, the time taken to caculate the NMR shielding tensors with the QM/MM model was approximately 1 hour, whereas the same calculation with full DFT took 12 hours using the same hardware.



\begin{table}
\caption{ $^1$H NMR chemical shifts $\delta$ (in ppm) for [Au$_{25}$(SCH$_3$)$_{12}$(SCH$_2$CH$_2$Ph)$_6$]$^{-}$ cluster.}
\label{tab:NMRH}
\centering
\begin{tabular}{llll}
\\ \hline\noalign{\smallskip}
Atom & QM/MM & DFT  & Experiment \cite{Gascon-Maran2011} \\
\noalign{\smallskip}\hline\noalign{\smallskip}
$\alpha$--CH$_{2, inner}$ & 2.92  & 3.23 & 3.80$^{broad}$ \\
$\alpha$--CH$_{2, outer}$ & 3.18  & 2.97 & 3.13\\  \\
$\beta$--CH$_{2, inner}$ & 2.93  &  2.95 & 3.13 \\
$\beta$--CH$_{2, outer}$ & 3.03  & 3.07 & t2.93\\ \\
m--CH$_{inner}$ & 7.28 & 7.24 & 7.15 \\
m--CH$_{outer}$ & 7.29 & 7.19 & 7.19 \\ \\
o--CH$_{inner}$ & 7.19 & 7.54 & 7.19  \\
o--CH$_{outer}$ & 7.23 & 7.35 & 7.14 \\ \\
p--CH$_{inner}$ & 7.23 & 7.02 & 7.08 \\
p--CH$_{outer}$ & 7.27 & 7.09 & 7.15 \\ \\
\hline
MUE (in ppm) & 0.18 & 0.18 & \\
\noalign{\smallskip}\hline
\end{tabular}
\end{table}

\section{Summary}

The use of X-ray crystallography to determine the structure of gold nanoparticles has been limited, owing partly to the difficulty of obtaining samples
of sufficiently uniform size for the growth of single crystals.
Until recently, there had been just one example of an Au cluster, Au$_{102}$(p-MBA)$_{44}$, which had been determined by crystallography \cite{RogerAu102}.
Since then, there have been successful crystal structure determinations of smaller sized gold clusters like Au$_{25}$ \cite{MurrayAu25} and Au$_{38}$ \cite{Qian2010}.
The recent availability of detailed structural data has created an opportunity for the construction and validation of computational methods for modeling
these systems.
Starting from the analysis of [Au$_{25}$(SCH$_3$)$_{18}$]$^{-}$ structure, we have developed generalized hybrid QM/MM models to accurately describe
the structure of monolayer protected gold clusters.
We have found that errors in the calculated geometries from our QM/MM model are $\sim$ 3\%, compared to the crystal structures.
In addition,
the isotropic chemical shifts obtained from NMR calculation agree very well among the QM/MM model, full DFT calculations and experiment.
Use of these hybrid models not only saves computational resources but also enables the study of physical properties and chemical reactions for larger clusters where physical insight can be obtained by describing only a small region of the system at a high QM level.

\section{Acknowledgements}
J.A.G thanks financial support from the Camille and Henry Dreyfus foundation, the NSF for a CAREER Award (CHE-0847340), and the Research Foundation at the University of Connecticut.
S.B would like to thank Prof. Robin C\^ot\'e at Dept. of Physics, University of Connecticut for helpful discussions and funding from US Department of Energy Office of Basic Sciences.

\newpage

\end{document}